# Effect of electron-electron interaction and disorder potential on the localization of two-dimensional electrons in a strong magnetic field

A.A. Vasilchenko

*National Research Tomsk State University, 634050 Tomsk, Russia*

The density functional theory is used to study the electronic structure of a quantum wire in a magnetic field at a filling factor $v \leq 1$. The Kohn-Sham equations are solved numerically for different values of electron densities and filling factors. The critical density, below which the electron density has a strong spatial inhomogeneity, was found in GaAs quantum wire. An empirical formula for the critical density is obtained. The effect of the disorder potential on the electron density profile in a two-dimensional electron gas is estimated.

**1. Introduction**

The electron-electron interaction plays an important role in such phenomena as Wigner crystallization, quantum Hall effect (QHE), conductance quantization, spontaneous spin polarization in zero magnetic field, metal-dielectric transition, phase separation in two-dimensional electron systems (see review [1]) and others. Many of these phenomena are still far from the complete theoretical explanation. Results of recent experimental [2-12] and theoretical works [13-15] indicate that our understanding of QHE is still very limited.

In Ref. [16], to explain the fractional QHE, it was proposed that two-dimensional electrons have a spatially inhomogeneous state, such as a Wigner crystal or a charge density wave. Experimental studies have shown the possibility of the formation of a phase-inhomogeneous state in two-dimensional systems in a magnetic field at filling factors $v = 1, 2, 1/3, 1/5$ [9–11] and $v < 0.5$ [12]. Based on the obtained experimental data, it was concluded in Refs. [9–11] that domains of a pinned Wigner crystal are formed near the filling factors $v = 1/3$ and $v = 1/5$. In the integer QHE regime, it was shown that there is a strong localization of charge carriers at $v = 1, 2$, and the formation of Wigner crystal domains is possible in the vicinity of these filling factors [9–11]. Experimental data [12] show that the formation of a two-phase inhomogeneous state as a mixture of Wigner crystals and electron liquid is possible in the MgZnO/ZnO electron system at $v < 0.5$.

The most impressive results on taking into account the electron-electron interaction in two-dimensional quantum dots were obtained in Refs. [13–15, 17-21], in which the many-particle Schrödinger equation was solved numerically. Results of numerical solution of the many-particle Schrödinger equation for the number of electrons $N < 10$ show that the energy spectrum of electrons has interesting features. In particular, the ground and metastable states of a electron system in a magnetic field are observed only at certain values of the total angular momentum of electrons. The results of calculations show that energy spectrum of electrons in quantum dots has a gap, whose origin is associated with electron-electron interactions. Exact calculations also show that the electron density distribution is spatially inhomogeneous and electrons crystallize and form a Wigner molecule at a filling factor $v < 0.4$.

In this work, we use the density functional theory taking into account the exchange energy in the local density approximation to study the electronic structure of a wide quantum wire in a magnetic field at a filling factor $v \leq 1$.

## 2. Theoretical model

We consider the two-dimensional electron gas in a planar quantum wire. The magnetic field is directed perpendicular to the quantum wire plane. Inside the quantum wire, electrons are confined by a positively charged background with a density $n_p$ ($n_p$ is nonzero at $|x| \leq a/2$, where $a$ is the width of the quantum wire).

According to the density functional theory, the total energy of a multielectron system is the functional of the electron density $n(x)$:

$$E[n] = T[n] + \frac{1}{2}\int V_H(x)(n(x) - n_p)dx + E_x[n], \qquad (1)$$

where $T[n]$ is the kinetic energy of noninteracting electrons in magnetic field $B$, which is given by the vector potential $A = (0, Bx, 0)$. The second term in expression (1) is the Coulomb energy.

For the exchange energy we use the local density approximation

$$E_x = \int \varepsilon_x(n)n(x)dx, \qquad (2)$$

where $\varepsilon_x = -\pi\sqrt{2\pi}Ln$, $L$ is the magnetic length.

We use the atomic system of units, in which the energy is expressed in units of $Ry = e^2/(2\varepsilon a_B)$, and the length in units of $a_B = \varepsilon\hbar^2/(m_e e^2)$, where $m_e$ is the effective electron mass, $\varepsilon$ is the dielectric constant. All calculations are performed for GaAs quantum wires, for which $\varepsilon = 12.4$ and $m_e = 0.067m_0$ ($m_0$ is the free electron mass). For GaAs we get $a_B = 9.8$ nm, $Ry = 5.9$ meV.

By minimizing functional (1) one obtains the Kohn-Sham equations:

$$-\frac{d^2\psi_k(x)}{dx^2} + \frac{(x-kL^2)^2}{L^4}\psi_k(x) + V_{eff}(x)\psi_k(x) = E_k\psi_k(x), \qquad (3)$$

where $V_{eff}(x) = V_H(x) + V_x(x)$, $V_H(x) = 4\int_{-\infty}^{\infty}(n_p - n(x_1))\ln|x-x_1|dx_1$,

$V_x(x) = \dfrac{d(\varepsilon_x(n)\, n)}{dn}$,

$n(x) = \int_{-k_F}^{k_F} \dfrac{dk}{2\pi}\psi_k^2(x)$, $k_F = \pi n_p a$.

The kinetic energy is given by

$$T = \int_{-k_F}^{k_F}\frac{dk}{2\pi}(E_k - \int V_{eff}(x)\psi_k^2(x)dx) \qquad (4)$$

The energy per electron is measured from the kinetic energy of non-interacting electrons $1/L^2$:

$$E_{1t} = \frac{E}{n_p a} - \frac{1}{L^2}. \tag{5}$$

For a homogeneous electron gas, the energy per electron is

$$E_{1h} = -\pi \sqrt{n_p \nu}, \tag{6}$$

where $\nu = n_p 2\pi L^2$

## 3. Numerical results and discussions

The nonlinear system of Kohn-Sham equations was solved numerically for various values of $n_p$ and $\nu$. Figure 1 shows the energy per electron as a function of the quantum wire width at $n_p = 2\times10^{11}$ cm$^{-2}$ and various values of $\nu$. At $\nu = 1$, the energy per electron $E_{1t}$ decreases monotonically with an increase in the quantum wire width and approaches the value of homogeneous electron gas energy (6) at high values of quantum wire width. Note that at $\nu = 1$ the energy $E_{1t}$ at $a_m = 100$ nm is higher than the energy $E_{1h}$ by 0.3 meV. For wide quantum wires at $\nu = 1$, the electron density is close to $n_p$ and differs from $n_p$ only near the quantum wire boundaries [23]. At $\nu = 0.8$, the $E_{1t}(a)$ curve has a weak minimum at $a_m \approx 60$ nm. At $\nu = 0.5$ and $\nu = 0.2$, the energy per electron has minima at $a_m \approx 31$ nm and $a_m \approx 24$ nm, respectively. The minimum energy per electron at $\nu = 0.8$, 0.5, and 0.2 is less than the energy $E_{1h}$ by 0.25 meV, 1.1 meV, and 1.5 meV, respectively. For wide quantum wires at $\nu \leq 0.8$, one should expect a breakdown of the two-dimensional gas into stripes with a width of about $a_m$.

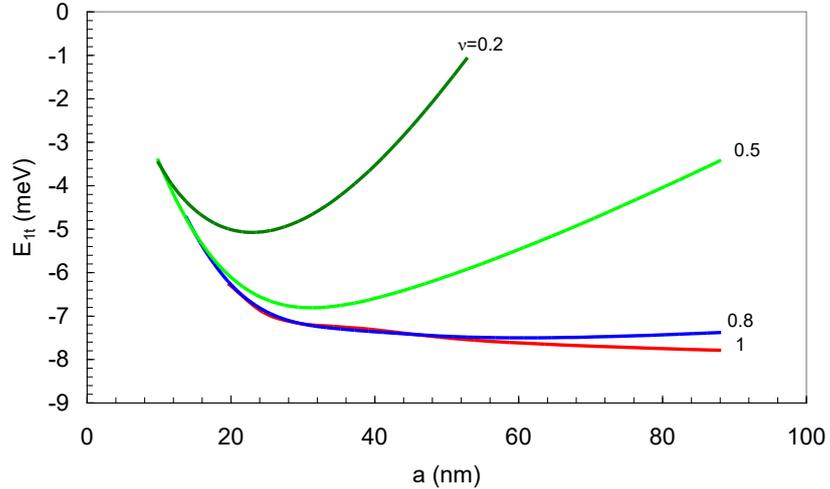

.
Fig. 1. Dependence of energy per electron on the width of the quantum wire at $n_p = 2\times10^{11}$ cm$^{-2}$.

Figure 2 shows the electron density profiles for different values of the filling factor. The selected values of the quantum wire widths at $\nu \leq 0.8$ correspond to the energy minima per electron (Fig. 1). As the magnetic field increases, the electron density profile narrows. At $\nu = 0.8$, the electron density at the quantum wire boundary is more than 10 times less than the density at the center, and at $\nu = 0.2$, the analogous value is almost equal to 500. Thus, we can assume that at $v \leq 0.8$, all electrons are localized inside the stripe.

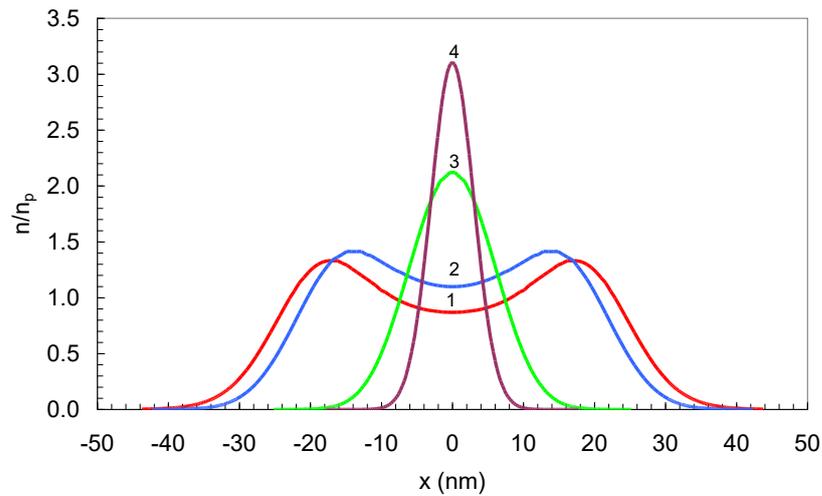

Fig.2. Electron density profiles at $n_p = 2\times 10^{11}$ cm$^{-2}$. Curve 1: $v = 1$, $a = 61$ nm; curve 2: $v = 0.8$, $a = 61$ nm; curve 3: $v = 0.5$, $a = 31$ nm; curve 4: $v = 0.2$, $a = 24$ nm.

At small quantum wire widths, the electron density at the boundary is comparable to the electron density inside the quantum wire (Fig. 3). As the quantum wire width increases, the electron density at the quantum wire boundary decreases. Note that for wide quantum wires near the center, the filling factor $v_e = n2\pi L^2$ is close to one for any value of $v$.

The calculation results at $v = 1$ showed that the highest localization of electrons occurs at $a \approx 30$ nm (the density at the quantum wire boundary is almost 10 times less than the density at the quantum wire center). As the width increases, the electron density at the boundary increases, and for wide quantum wire, the electron density is close to $n_p$. Therefore, at $v = 1$, electrons can be localized inside the quantum wire only under the influence of a disorder potential.

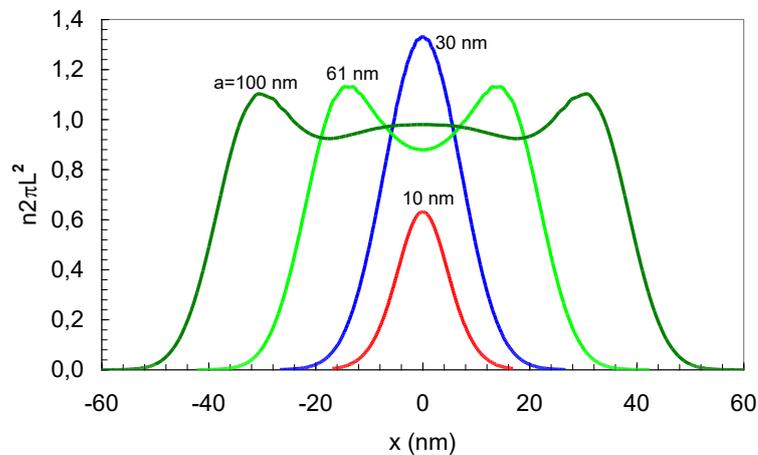

Fig. 3. Electron density profiles at $n_p = 2\times 10^{11}$ cm$^{-2}$ and $v = 0.8$

Let us estimate the effect of the disorder potential on the electron density profile in a two-dimensional electron gas. For a weak disorder potential $V_d(x)$, the energy per electron is as follows:

$$E_{1d} \approx E_{1t} + \frac{\int V_d(x) n(x) dx}{n_p a}. \qquad (7)$$

As an example, we consider the transition from the ground (localized) state to a state with a nonzero electron density at the quantum wire boundaries at $v \leq 0.8$. As seen from fig. 1, the difference in energy $\Delta E$ between the ground state and the nonlocalized state ($a \approx 20$ nm at $v = 0.8, 0.5$ and $a \approx 10$ nm at $v = 0.2$) is $1 \div 2$ meV. A disorder potential with an amplitude of the order of $\Delta E$ and a spatial scale of the order of several $a_B$ can lead to the formation of connected spatial regions of the two-dimensional electron gas.

As the electron density decreases, the role of the exchange interaction increases. At $n_p = 10^{11}$ cm$^{-2}$ and $v = 1$, a minimum appears in the dependence $E_{1t}(a)$ [22]. The energy value at the minimum is greater than the energy $E_{1h}$, and the electron density distribution remains spatially homogeneous for wide quantum wires. The calculation results show that at $n_p = 9\times 10^{10}$ cm$^{-2}$ the value of the energy minimum $E_{1t}(a_m)$ becomes less than $E_{1h}$ and the electron density distribution becomes highly inhomogeneous for wide quantum wires.

Figure 4 shows the dependence of the energy per electron on the quantum wire width at $n_p = 5\times 10^{10}$ cm$^{-2}$ and different values of $v$. The positions of the energy minima at $v = 1$ and $v = 0.8$ are close to each other, and as the magnetic field increases, the energy minima shift towards smaller quantum wire widths. At $v = 1$, the energy per electron (Fig. 4) is significantly less (by 2.8 meV) than the energy of a homogeneous electron gas $E_{1h}$, and at $v = 0.8, 0.5,$ and $0.2$, the difference in energies $E_{1h} - E_{1t}$ is 1.2 meV, 0.7 meV, and 0.8 meV, respectively.

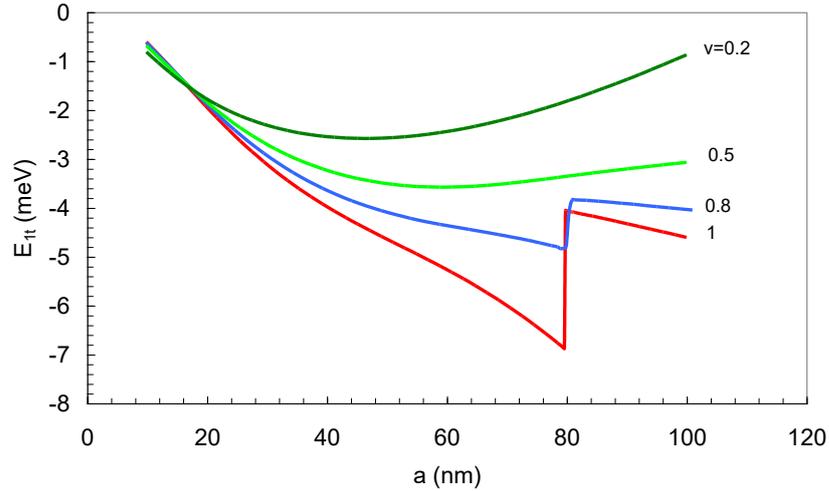

Fig. 4. Dependence of energy per electron on the width of the quantum wire at $n_p = 5\times 10^{10}$ cm$^{-2}$

The electron density changes periodically with a period of about 60 nm at $v = 1$ and $a \gg a_m$. Note that at $n_p = 7\times 10^{10}$ cm$^{-2}$, the energy per electron has a minimum at $a_m \approx 60$ nm, and the period is approximately equal to 50 nm for wide quantum wire [22]. By analogy with these results, we can assume that the period of change in the electron density at $v < 1$ will also be close to $a_m$. To determine the period at $v < 1$ more accurately, it is necessary to carry out calculations for wide quantum wire with different sets of occupied and empty states.

Figure 5 shows the electron density profiles at $n_p = 5 \times 10^{10}$ cm$^{-2}$ and quantum wire widths at which the energy per electron has a minimum. For any filling factors, the electrons are localized inside the quantum wire. In contrast to the case of high density, the electron density profile broadens with increasing magnetic field. This phenomenon is due to the increasing role of exchange interaction.. At $n_p = 5 \times 10^{10}$ cm$^{-2}$ and $v = 1$, the contribution of the exchange energy is highest, and with an increase in the magnetic field (decrease in $v$), the exchange energy decreases and at $v = 0.5$ it is compared with the kinetic energy. In experimental works [9–11], a strong localization of charge carriers at $v = 1$ was also obtained.

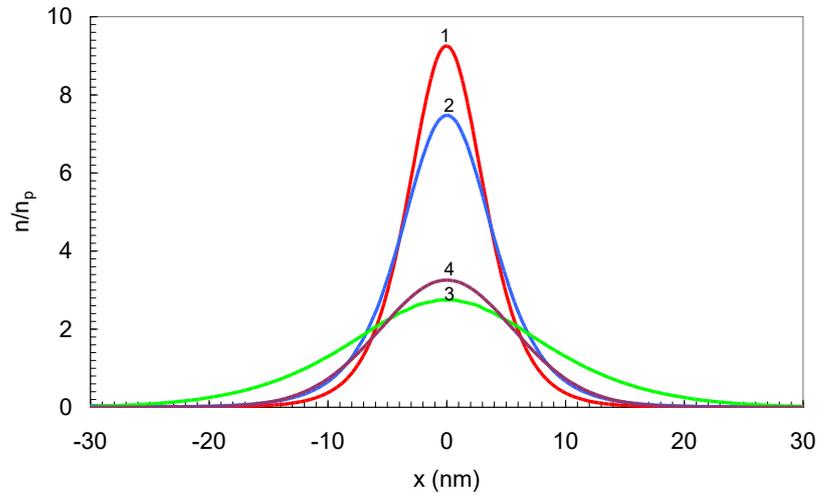

Fig. 5. Electron density profiles at $n_p = 5 \times 10^{10}$ cm$^{-2}$. Curve 1: $v = 1$, $a = 79$ nm; curve 2: $v = 0.8$, $a = 79$ nm; curve 3: $v = 0.5$, $a = 59$ nm; curve 4: $v = 0.2$, $a = 47$ nm.

Let us estimate the value of the critical electron density. At high electron densities or small filling factors, the kinetic energy of a free electron significantly exceeds the exchange potential $V_x(n_p)$ and the value of $V_x(n_p)$ can be considered as a perturbation. The calculation results show that the transition to an inhomogeneous density occurs at an exchange potential $V_x(n_p)$ approximately three times higher than the kinetic energy of a non-interacting electron. In this case, we obtain the expression $n_c \approx 0.1\ v^3/a_B^2$ for the critical density. The results of the calculations show that the kinetic energy differs greatly from the kinetic energy of noninteracting electrons at an average electron density $n \leq n_c$. For example, at $n_p = 5 \times 10^{10}$ cm$^{-2}$ and $v = 0.8$, the kinetic energy increases by about four times compared to the kinetic energy of a free electron, and at $v = 0.5$ and $v = 0.2$, the difference in kinetic energy is 11 percent and 1 percent, respectively .

As in the case of $n_p = 2 \times 10^{11}$ cm$^{-2}$, the states at $n_p = 5 \times 10^{10}$ cm$^{-2}$ and $a = 10 \div 20$ nm are non-localized and the disorder potential can lead to the formation of connected spatial regions. As $n_p$ decreases, the energy difference $\Delta E$ between the ground state and the nonlocalized state increases, and the value of $\Delta E$ changes from 2 meV to 5 meV at $n_p = 5 \times 10^{10}$ cm$^{-2}$ (Fig. 4). Note that the sum of the exchange and Coulomb energies is almost independent of the magnetic field at $a = 10 \div 20$ nm (Figs. 1 and 4).

The chosen symmetry can be broken and a spatially inhomogeneous distribution of the electron density along the stripe is possible. In this case, electron drops with a finite number of electrons can be formed. The transverse dimensions of the drop are comparable to the width of the quantum well; therefore, to confirm this hypothesis, it is necessary to perform calculations for a three-dimensional anisotropic drop. In the case of the formation of electron drops, the width of the Hall plateau at $\nu = 1$ will increase with a decrease in the number of electrons in the droplet [21]. A drop model of the quantum Hall effect in the framework of disk geometry was proposed in Ref. [23]. In the drop model, the transition to a new Hall state occurs when the number of electrons in the drop changes.

## 4. Conclusion

The Kohn-Sham equations for electrons in a magnetic field are numerically solved. It is found that in GaAs quantum wire at densities $n \leq 9 \times 10^{10}$ cm$^{-2}$, the distribution of electron density has a strong inhomogeneity and the electron gas consists of unconnected stripes. At high densities, the electron density distribution will be spatially homogeneous at $\nu = 1$ and inhomogeneous at $\nu < 0.8$. A disorder potential with an amplitude of several meV (with a decrease in the average electron density, the amplitude increases) and a scale on the order of several $a_B$ can lead to the formation of both connected and unconnected regions of a two-dimensional electron gas. Note that experimental studies have shown the possibility of the formation of a phase-inhomogeneous state in two-dimensional systems at $\nu = 1$ [9–11] and $\nu < 0.5$ [9–12].

**Acknowledgements**

This work was supported by the State Assignment of the Ministry of Science and Higher Education of the Russian Federation (project No. FSWM-2020-0048).